# Cognitive Modelling Aspects of Neurodevelopmental Disorders Using Standard and Oscillating Neighbourhood SOM Neural Networks[#]

**Spyridon Revithis**[1], **Nadine Marcus**[2]

[1,2]School of Computer Science and Engineering
University of New South Wales, NSW Sydney, Australia

[1]revithiss@cse.unsw.edu.au, [2]nadinem@unsw.edu.au

**Abstract**

**Background/Introduction**: In this paper, the neural network class of Self-Organising Maps (SOMs) is investigated in terms of its theoretical and applied validity for cognitive modelling, particularly of neurodevelopmental disorders.

**Methods**: A modified SOM network type, with increased biological plausibility, incorporating a type of cortical columnar oscillation in the form of an oscillating Topological Neighbourhood (TN), is introduced and applied alongside the standard SOM. Aspects of two neurodevelopmental disorders, autism and schizophrenia, are modelled using SOM networks, based on existing neurocomputational theories. Both standard and oscillating-TN SOM training is employed with targeted modifications in the TN width function. Computer simulations are conducted using revised versions of a previously introduced model (IPSOM) based on a new modelling hypothesis.

**Results/Conclusions**: The results demonstrate that there is strong similarity between standard and oscillating-TN SOM modelling in terms of map formation behaviour, output and structure, while the oscillating version offers a more realistic computational analogue of brain function. Neuroscientific and computational arguments are presented to validate the proposed SOM modification within a cognitive modelling framework.

**Acknowledgments**: We would like to thank Dr. William H. Wilson for his insightful discussions and very valuable feedback to an earlier version of this work.

---

[#] This paper is a substantially revised and expanded version of [114].

# 1. Introduction

This paper investigates mental disorders using biologically relevant computational models. The first of Kandel's five principles that constitute his "common framework for psychiatric and neural sciences" outlines the biological (brain-related) nature of all mental processes, and describes the origin of "behavioural disorders that characterise psychiatric illness" (even when caused by the environment) as "disturbances of brain function" [1]. Furthermore, there is a continuous trend in the mind sciences towards interdisciplinary research, where each discipline (e.g. neuroscience, psychology, computational modelling) facilitates the work of others while providing traceable causation between the different levels of inquiry [2].

Cognitive phenomena involving processes, behaviour and biological determinants have long been studied using computational models; the latter approach constitutes a valid framework with demonstrated applicability, offering diverse methodological tools, clarity, scalability and accuracy [3][4]. While mathematical and statistical models describe a phenomenon in an abstract and static fashion without reproducing it, computational models can be executed and observed; this offers the opportunity to empirically study the model's behaviour (from input to output) in a constructive manner alongside the phenomenon it models, and benefit from the emergence of unforeseen new ways of understanding and evaluating it [5]. Numerous modelling studies, employing artificial neural networks, focus on cognition and development [6][7][8][9][10][11][12][13]; on the latter, Shultz [3] evaluates a range of neural networks models. McClelland et al. [14] considers the connectionist approach to be insightful in relation to the actual mechanisms behind cognitive phenomena and biologically relevant in understanding cognition and development compared to probabilistic approaches. A 2015 classification places the connectionist approach in close proximity to nature-inspired computing [15].

The class of SOM (Self-Organising Map) neural networks is of particular significance; a SOM utilises unsupervised Hebbian-style [16] machine learning and presents a comparable computational analogue to topographic brain cortical maps in terms of structure and output [17][18][19][20][21][22]. Von der Malsburg discusses the value of SOM neural networks in overcoming theoretical challenges regarding mental phenomena and their computational analogues [23], as well as the importance of self-organised network patterns ("net fragments") in natural intelligence [24]. Mareschal and Thomas [25] discuss the importance of self-organisation in cognition and cognitive development, and present a range of relevant connectionist models including Cohen's [26] and Gustafsson's [27] models of autism.

Of the six principles for neurocomputational models of cortical cognition (biological realism, distributed representations, inhibitory competition, interactivity/recurrence, error-driven task learning, and Hebbian model learning) discussed by O'Reilly [28], all except error-driven learning are supported by self-organising feature maps. However, there seems to be no strong evidence in the research literature of error-driven learning in primary cortical map organisation. The work presented here contributes towards strengthening the first principle (biological plausibility) and is distinct from other research that employs computational methods (e.g., [29][30]) lacking any explicit links to either neurocognitive mechanisms or biological processes.

SOM networks were first proposed by Willshaw and von der Malsburg [31] in relation to the retinotopic-mapping problem. However, the Kohonen SOM [32][33] offers a more straightforward computational implementation while presenting with important neurobiological and statistical properties (input-space approximation, topological ordering, density matching and feature selection, as well as non-linear representations and data compression) that are relevant and particularly useful to the study and modelling of cortical maps in the brain [34].

This paper presents SOM cognitive modelling aspects of neurodevelopmental disorders. Section 2 introduces a SOM modelling framework and a novel modification in SOM training with cognitive modelling implications in terms of biological plausibility. In the next two sections (Sections 3 and 4), two neurodevelopmental disorders (autism and schizophrenia) are being investigated and certain characteristics modelled using two closely related SOM models. Finally, there is a discussion (Section 5) on the modelling specifications, as well as a summary and considerations for future work.

## 2. The SOM Modelling Framework

### 2.1 Computational and Cognitive Modelling Aspects of SOM Neural Networks

A SOM neural network has a dual-layer structure that performs non-linear unsupervised machine learning. It can map its input layer, which is directly linked to its situated environment as a series of n-dimensional input patterns, onto a lattice (the 'Kohonen layer'), usually 1- or 2-dimensional; the output (Kohonen) layer can eventually form a representation of the pattern structure of the input-space. The representation of environmental input in the output layer (the trained map) is performed in a topologically ordered fashion, maintaining the non-linear input data distribution, and it involves dimensionality reduction since the map's dimensionality is typically smaller than that of the input space [34]. Figure 1 shows an abstract depiction of a two-dimensional SOM; each input-layer pattern vector connects to each neuron on the map (this is not shown in full detail in order to simplify the image). Note the identical dimensionality of the weight vectors for both the map neurons (in grey) and the input patterns.

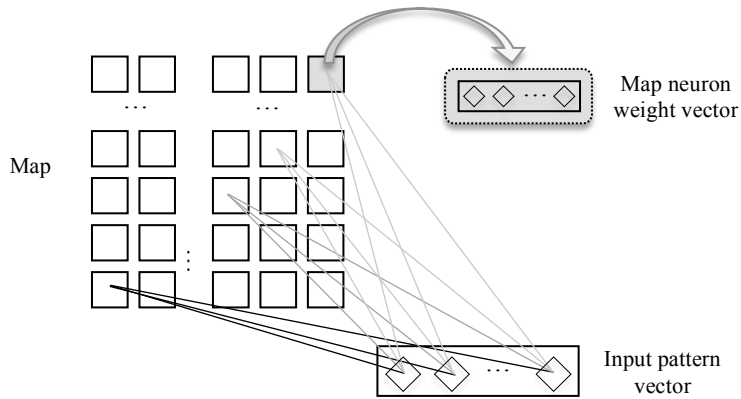

**Figure 1: A two-dimensional SOM depicting some indicative connections to the input layer.**

The training of a SOM, resulting in a formed map, consists of four parts [34]: initialisation of map synaptic weights; competition between map neurons; cooperation between map neurons; and synaptic weight adaptation. This process is briefly and selectively reviewed here, and discussed from a cognitive modelling standpoint in order to provide the necessary context for the modifications introduced in the next section. The latter three parts of the SOM training process (competition, cooperation, and adaptation) are repeated in a loop for a certain number of epochs (algorithmic iterations). In each epoch, all input patterns are presented and weights adjusted; this is repeated until the weights converge. Specifically, in the competition phase, an input-space vector is presented and a winning neuron is determined, based on which neuron's weight vector is closest (i.e., most similar) to the input vector. In the cooperation phase, the winning neuron becomes the centre of a cooperative process extending over an area according to a Topological Neighbourhood (TN) function. In the synaptic adaptation phase, Hebbian learning takes place, where the weights of all the neurons inside the winning neuron's TN area in the map are adjusted to become more similar (closer) to those of the input vector by an amount determined by three factors: their lateral distance to the winning neuron; a TN width function σ; and a learning rate function η that has an exponential decay.

The neurobiological relevance of the implementation of the SOM formation process in terms of the neuronal lateral activation and inhibition mechanisms has significant cognitive modelling implications. The standard SOM formation algorithm [34] employs a Gaussian TN function that is spatially symmetric around the winning neuron, and its value decreases with the lateral distance to the winning neuron, but the TN width σ(*n*) decreases exponentially with the epoch number *n*, as indicated in Figure 2.

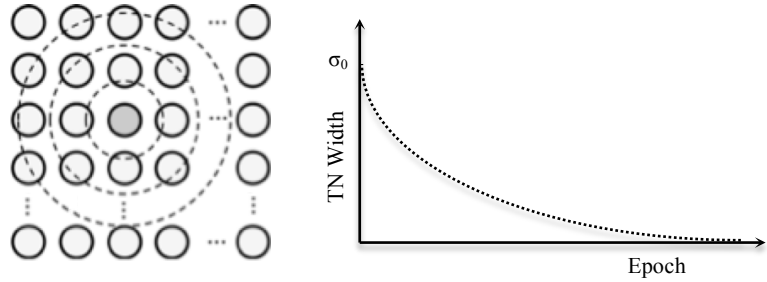

**Figure 2: Decreasing TN width around a winning neuron in a two-dimensional SOM during training. The dark grey neuron serves as an example of a neuron that wins the competition a certain number of times during training; each time the TN (depicted as decreasing concentric circles) is smaller than before.**

Mathematically, the TN width function σ(*n*) can be expressed as

$$\sigma(n) = \sigma_0 \cdot \exp\left(-\frac{n}{\tau_1}\right), \quad n = 1, 2, \ldots, t$$

where $\sigma_0$ is the initial TN width, $\tau_1$ is a time constant, *t* is the total number of epochs, and *n* is the current epoch. The learning rate function η(*n*) has identical structure to σ(*n*), using corresponding $\eta_0$ (initial earning rate) and $\tau_2$ (time constant), but, unlike σ(*n*), it is not part of the TN function:

$$\eta(n) = \eta_0 \cdot \exp\left(-\frac{n}{\tau_2}\right), \quad n = 1, 2, \ldots, t.$$

The TN function *h(n)* for a given winning neuron *i(x)* and an excited SOM neuron *j* can be expressed by the formula

$$h_{j,i(x)}(n) = \exp\left(-\frac{d_{j,i(x)}^2}{2 \cdot \sigma^2(n)}\right), \quad n = 1, 2, \ldots, t$$

where *x* is the current input neuron (weight vector), and *d* is the lateral distance between the two neurons (*i(x)* and *j*). Then, the weight update function for weight vector $w_j(n)$ for a given neuron *j* within the TN can be expressed as

$$w_j(n+1) = w_j(n) + \eta(n) \cdot h_{j,i(x)}(n) \cdot [x(n) - w_j(n)], \quad n = 1, 2, \ldots, t$$

where *x(n)* is the current input vector at time step *n* and *i(x)* is the winning neuron for a given *n*.

Experimentation with the η and σ parameters verifies that the SOM algorithm works reasonably well even when η is held constant at not too high a value, which is not surprising since learning rate generally does not influence the selection area of SOM neuronal activation (see also [35]), and that a decreasing σ is essential for map formation. This constrains the form of the modifications to the SOM algorithm described in the next section.

The SOM algorithm as just presented produces a neurally realistic outcome - an analogue of a somatotopic cortical map - but the fact that η and σ decay exponentially towards zero means that SOM learning stops when these parameters approach zero. This is not realistic - somatotopic maps do not freeze, as demonstrated by the classic work of Merzenich et al. [36][37]. In their work, a part of the somatotopic map for an adult monkey's digits was mapped using single-neuron electrode recording, and then one of the monkey's middle digits was cut off. Over a number of weeks, they observed that the cortical region previously sensitive to touching the digit that they had removed became progressively sensitive to touching the digits on either side. This demonstrated continued learning/re-learning in the somatotopic map. Brain neuroplasticity, in general, supports the existence of cortical maps that are adaptable and ever changing [38].

A key correspondence between the relevant neurobiological and computational studies is based on the vertical column hypothesis of the neocortex [39][40][41][42]. It is not entirely clear whether neural columns are anatomically fixed or an emergent functional aspect of the neocortex [38][42][43]. Although it has been a matter of debate since first proposed, the columnar organisation of the cortex remains the most widely adopted hypothesis for cortical processing of information [44][45]. SOM networks natively model principal functional and anatomical aspects of neuron columnar ensembles in the cortex, and a significant role is played by the TN component of the network. Specifically, a neural column can be modelled, in principle, by the group of neurons within a SOM TN [27][38][46]. In this way, a trained SOM corresponds to a macrocolumn and includes a number of minicolumns.

The representational structure of a SOM is largely the result of its adaptation process in relation to the input space; this process is dictated by synaptic weight adjustments surrounding winning neurons and corresponds to the biological mechanisms of neuronal excitation and lateral inhibition [47][33]. Lateral inhibition is considered a key mechanism in the human cortex [48]. Structural changes in the cortex are driven by consistent and reliable patterned sensory stimulation and are regulated by a balanced combination of neural excitation and inhibition; intracortical inhibition plays a significant developmental role [49]. Research on brain plasticity, centering around the role of excitation-inhibition balance, could prove catalytic in therapeutic interventions for neurodevelopment disorders such as autism and schizophrenia [50][51][52].

Existing SOM modelling of brain disorders has underscored the significance of TN width as an implementation of lateral inhibition with neurobiological plausibility. The TN can be regarded as the "source of power" [53] in these models. It has been argued that excessive lateral inhibition, linked to narrow neural columns, can decisively contribute to the development of autistic traits [27]. High neuromodulator activity, associated with acute delusional states, has also been linked to excessive lateral inhibition [38]. These two neurocomputational theories will be discussed in more detail later. Furthermore, working memory capacity seems to be limited by lateral inhibition [54] while working memory impairment is considered an important area of cognitive impairment in schizophrenia [55][56][57]. Lanillos et al. review the different connectionist approaches to the modelling of autism and schizophrenia while noting the predominant focus on perceptual defects (autism) and hallucinations & delusions (schizophrenia) as well as the utilisation of the neural inhibition/excitation mechanism [58].

**2.2 SOM TN Width Oscillation**

During SOM formation there are extensive and repeated local interactions between map neurons that follow a Hebbian principle of neuronal activation, complemented by a lateral inhibitory type of regulatory mechanism. It should be noted that local and long-range neuronal communication, as well as neural oscillation, are integral functions of the brain; oscillatory activity linked to neuronal inhibition is claimed to be of great significance to neural information processing and presented in numerous studies of brain disorders including schizophrenia and autism [59][60][61][62][63].

The cited research that was outlined here and in the preceding section indicate a close link between lateral inhibition, neural oscillation and cortical organisation, and support a claim for the cognitive modelling significance of the SOM TN width in biological neural networks. Accordingly, this paper introduces a modification in the SOM cooperative phase that aims to address the biological implausibility of having SOMs learn once (over a number of time steps) and then freeze the weights, especially in modelling brain disorders.

In particular, the standard SOM TN width function, which is a component of the overall SOM TN function, can be substituted by a function that produces numerous local temporal decreases in its value over a single SOM formation session instead of a single global decrease in value. This will result in a type of TN width oscillation during SOM training, consisting of a series of temporally shortened and concatenated standard TN width functions; hence, in a single SOM training session multiple piecewise exponentially decreasing segments will be realised, as illustrated in Figure 3.

The new SOM TN width function can be expressed by the formula

$$\sigma'(n) = \sigma_0 \cdot \exp\left(-\frac{(n+1) \bmod t'}{\tau'_1}\right), \quad n = 0, 1, 2, \ldots, t-1$$

where $\sigma_0$ is the initial TN width, $\tau'_1$ is a time constant, and $n$ is the current epoch. The constant $t' = t/c$, where $c$ is the oscillation constant, determines how many times the TN width will reset to $\sigma_0$ and start decreasing again during the t processing steps.

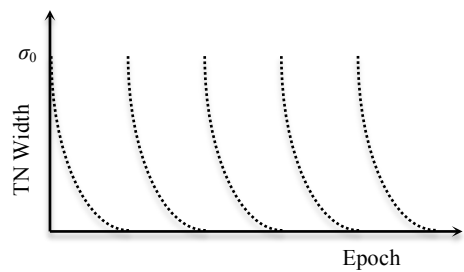

**Figure 3: SOM oscillating TN width; see Figure 2 for a comparison.**

Oscillation is necessary in a biologically plausible model, otherwise learning would essentially cease forever when the TN width approached zero. Depending on the temporal modelling hypothesis, a single SOM formation session could be associated with a single cognitive task, a single day, or even the duration of life, also depending on whether the trained SOM corresponds to a low-level or a high-level cortical map in terms of knowledge representation and function. Irrespective of the temporal modelling hypothesis, however, standard SOM formation, which is characterised by single descent TN width, neither can sustain brain-like plasticity properties nor protect from sensory desensitisation (i.e., lack of neural reorganisation based on sensory input); this is true also in the case where the overall duration of the SOM formation process is greatly increased. On the other hand, an oscillating TN width is more effective in supporting SOM continuous learning, the preservation of SOM topological ordering, and better responding to input stimuli, all of which is vital to the reorganisation process. It is reasonable to assume that cortical map formation is continuous throughout life [38], but not observed a lot because observation involves drastic investigations and, mostly, not much actually changes in terms of stimulation patterns. It becomes apparent that, unlike a standard SOM, an oscillating TN SOM is a mechanism that is a step in the right direction in addressing the reality of cortical map changes from a cognitive modelling standpoint.

In neural oscillation, synchrony can be considered at various levels. One physiological interpretation of the proposed TN-oscillating SOM would highlight the analogy between an oscillating local field potential, representing the combined activity of a local population of neurons, and a SOM oscillating TN width session, involving the cooperative activity of the population of neurons within the TN. Again, such rhythmic patterns of activity can be identified and compared at either the level of individual neurons or at a macroscopic level of neuron ensembles.

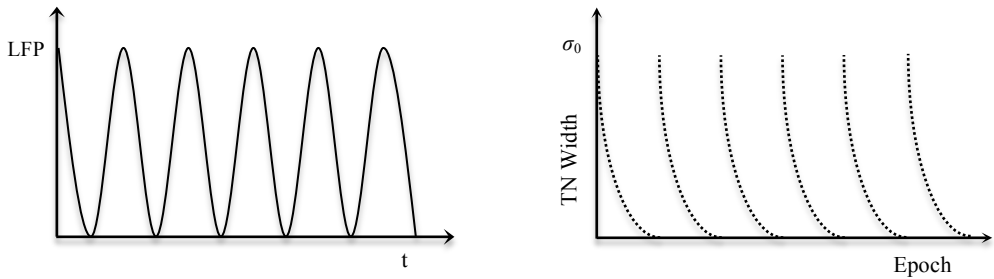

Figure 4: Neural and TN-width SOM oscillation.

As an example, in Figure 4, the local field potential (LFP) of a group of neurons is illustrated on the left abstract graph, representing the dominant activity in a given cortical area (examples of this graph based on specific data can be found in the literature; see, for example, [62]). On the right graph, the collective activity of SOM neurons within a given TN is implicitly depicted; the larger the TN width, the larger the number of SOM neurons that have their weights adjusted. In both cases, LFP and TN width are population measures; the oscillation observed is macroscopic. While synchrony is a property relating to spike-train signalling in the brain, a way of describing brain activity that is different to the rate-based models here, a broad analogy between TN width oscillation in the model and synchronous oscillation in the brain can still be insightful.

There are also arguments to support a hypothesis that SOM stability or entrenchment during map formation associates with neural oscillatory sensory prediction; entrenchment can be characterised as the formation of statistically and structurally dominant patterns with temporal sensitivity in terms of map-training order [64]. Firstly, meaningful and important stimuli are often temporally regular [65][66]; due to the mechanism of SOM formation, statistically dominant input patterns presented in a rhythmic fashion will enhance existing SOM representations. Secondly, sensory synchrony (i.e., what pattern appears when), during SOM training, can determine the map's representational stability and temporal predictability of its environment. Lastly, there are studies on cortical oscillation [65][66] that offer neurobiological explanations of how the brain utilises low-frequency/low-level oscillatory mechanisms for sensory prediction and selection.

**2.3 The IPSOM Model**

IPSOM (Interlocking Puzzle SOM) is a novel computational SOM spatial behavioural model of the sequences in which humans complete interlocking puzzles [67]. It employs a complex weight-encoding scheme that is based on a custom spatial position matrix, and uses a pseudo-random generator to select input-space instances; the latter are drawn from a predesigned synthetic training set of four patterns, corresponding to easily recognisable real-life puzzle completion strategies. Training IPSOM using a representative set of puzzle completion sessions (consisting of an adequately repeated presentation of the training patterns) results in a behavioural map containing statistically dominant patterns (strategies) of puzzle completion.

An IPSOM case study model, implemented on a 6x6 output lattice (map), was evaluated previously for the case of 4x5 puzzles (20-piece rectangular puzzles) with a computer-simulated group of people. Each virtual person was set to use one of the four available predetermined puzzle completion strategies per puzzle completion session, without exhibiting selection bias. The designed set of strategies is illustrated in Figure 5.

Each of the four predesigned puzzle completion patterns (H, V, PH, PV) is depicted in Figure 5 by a radar-graph: the radial axis scale corresponds to the range of encoded position values on the puzzle board (each puzzle piece is identified by a numeric position value), and the angular axis scale corresponds to the puzzle completion sequence range ($1^{st}$ puzzle piece placed, $2^{nd}$ puzzle piece placed, ..., $20^{th}$ puzzle piece placed on the board). Both the numeric position values and the sequence numbers of all the puzzle pieces were encoded in the weight vector of each IPSOM neuron. Once all the points on the radar-graph have been connected, a distinct visualisation of the puzzle completion pattern appears; in these graphs, even slight pattern deviations (e.g., very similar, but not identical, puzzle completion strategies) can be distinguished. Paired with each graph, a puzzle board image illustrates the puzzle completion order in a more conventional visual manner. The guiding principles in the design of the specific training set of puzzle completion patterns were: real-life, uncomplicated strategies; utilisation of topological clustering; and emphasis on the puzzle board periphery. The evaluation showed that IPSOM was an efficient model of the behavioural domain [67].

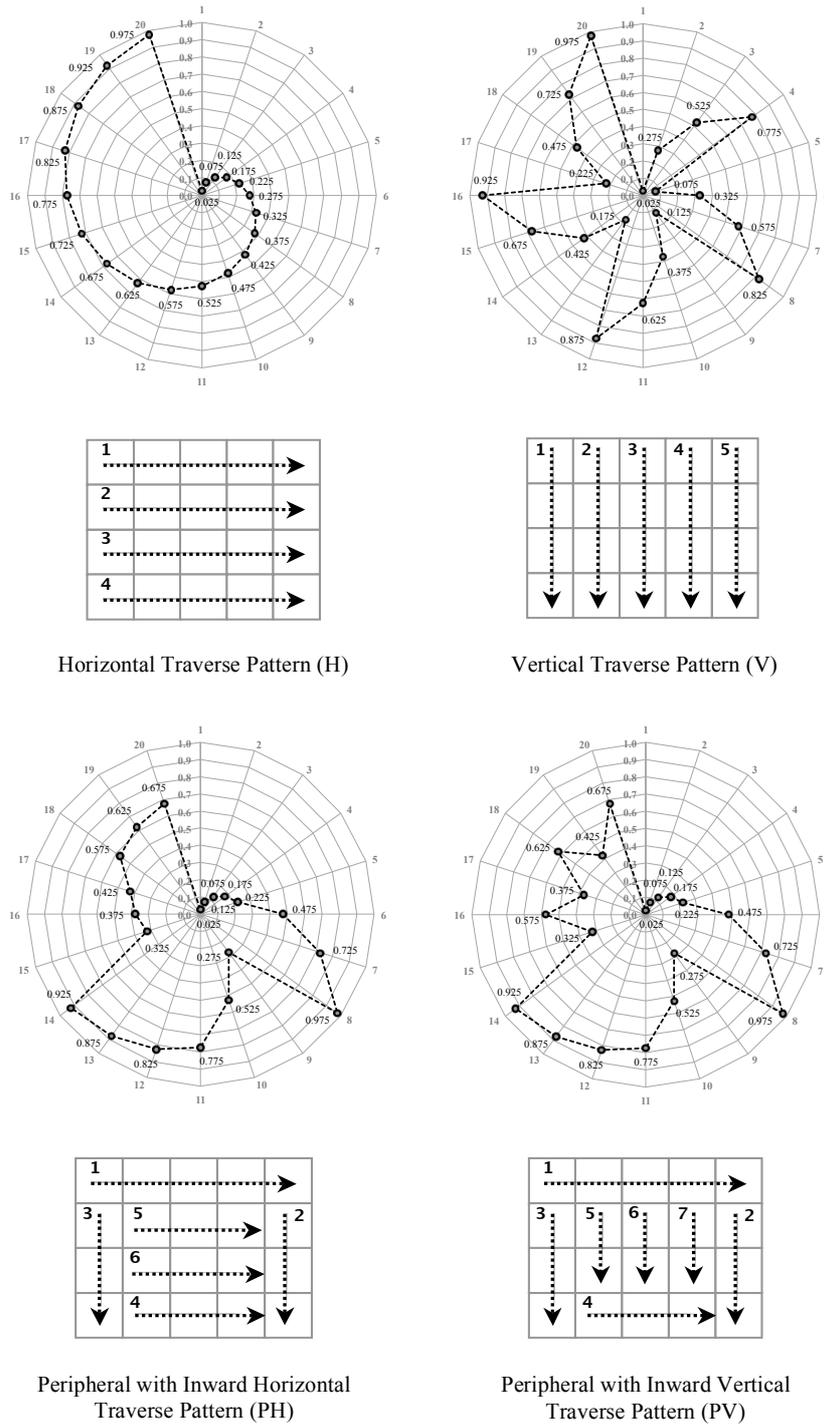

**Figure 5: IPSOM training set patterns (strategies). The radial graphs enable visualisation of the patterns, based on a puzzle-piece position (0.025, 0.075, …, 0.975) and completion-order (1st, 2nd, 3rd, … 20th) encoding, in a way that the conventional illustration of the accompanying puzzle boards do not.**

In the work presented here, two computationally related models based on IPSOM (IPSOM-A and IPSOM-D) are employed as test-beds for cortical map spatial perception. The working hypothesis is that IPSOM is not only a behavioural model, as shown previously, but also a cognitive model of how humans perceive and internalise puzzle completion strategies when they are presented with puzzle completion sessions/examples. It is assumed that an average person would form an internal representation of the dominant strategies; a cortical map would retain the domain specific knowledge, modelled by a trained SOM. The expectation is that IPSOM would represent the training patterns from the puzzle completion sessions in a topologically ordered fashion, where neighbouring patterns are also visually similar (Figure 6).

|   | 1 | 2 | 3 | 4 | 5 | 6 |
|---|---|---|---|---|---|---|
| 1 | **PV** | PV | PV- | V- | V | **V** |
| 2 | PV | PV- | PV- | V- | V- | V |
| 3 | PV- | PV- | ~ | ~ | V- | V |
| 4 | PH- | PH- | ~ | ~ | H- | H- |
| 5 | PH | PH- | PH- | H- | H- | H- |
| 6 | **PH** | PH | PH- | H- | H | **H** |

**X**    Primary core neuron (Optimal pattern representation)
X    Core neuron (Good pattern representation)
X-    Weak neuron (Poor pattern representation)
~    Undecided neuron (Excessively transitional pattern)

**Figure 6: An illustrated example of a trained 6x6 IPSOM.**

The working hypothesis introduced here serves both as a tool for scaling-up SOM cognitive modelling and as the modelling goal itself. The fact alone that IPSOM models a domain that is not typically found in neuropsychological or psychiatric studies, including ones on autism and schizophrenia, does not detract from its applicability to these studies, especially from a cognitive-behavioural standpoint where the focus is on map formation characteristics. Furthermore, due to its complex weight-encoding, IPSOM is a good alternative candidate to validate SOM modelling in addition to the more abstract computer simulations, and the ones focusing at the lower biological level, employed in cognitive modelling studies of brain disorders.

The fact that the number of IPSOM map nodes (36 neurons) is significantly larger than the set of training patterns (four predesigned strategies) allows for the recording of extra information about the model's representational behaviour. Customarily, SOM models are looking for clusters in large amounts of data, so having fewer map nodes than input patterns is appropriate, but our purposes are different. Specifically, it allows us to observe the pattern placement on the map, or lack thereof; for instance, when no training bias is applied, a typical map will tend to accommodate equidistant patterns, as shown in Figure 6, in accordance with SOM properties, especially density matching and topological ordering. Also, it allows us to record the transition between well-represented training set patterns, and the existence and characteristics of atypical patterns on the map. As we will discuss next, such data are potentially strongly indicative of certain cognitive and behavioural tendencies when using the model to study aspects of neurodevelopmental disorders including autism and delusions in schizophrenia.

# 3. The IPSOM-A Model

## 3.1 A Neurocomputational Approach to Autism

Autism, a pervasive developmental disorder (PDD), has been the subject of interdisciplinary research since Kanner's [68] and Asperger's [69] papers. Its aetiology remains unknown, but it is considered to be neurobiological in nature [70]. Unfortunately, the existing diagnostic & classification manuals, DSM and ICD [71][72][73][74], prescribe a socio-psychological behavioural approach that lacks aetiological considerations.

Autism is characterised by atypical perception and its internal representation. Sensory input largely fails to integrate into the existing memory (schemas) due to impairment in the abstraction process [75]. People with autism have difficulty in distinguishing salient features among details that are of less importance [76]. Internal manipulation of information is also problematic for them, where it appears that central executive control is necessary [77].

A neural circuit theory of autism, proposed by Gustafsson [27], is based on the aforementioned concepts about autistic perception. Gustafsson has developed a neurobiological explanation for the lack of drive for central coherence, a key aspect of behaviour in autistic people [78]; he postulates that deficiencies in the formation of brain cortical maps result in autistic attributes and that these maps are characterised by narrow neural columns. This affects the mechanism of memory formation and retrieval, causing failure in feature extraction, since "autistic raw data memory" (i.e., storage of raw sensory data) operates in place of "feature memory" due to "inadequate cortical feature maps". Raw-data memory is intrinsically linked to the diagnostic criteria for autism at the behavioural level [27]. Cortical maps in autistic people have deficiencies in feature distinction and preservation, and lack the capacity of effective internal representations of salient perceptual data; this results in a raw-data memory that lacks refined representations and effective generalisations of environmental stimuli. Gustafsson's columnar hypothesis for autism has been supported by neuroanatomical imaging data [79] showing cortical minicolumnar abnormalities (narrow and many minicolumns) in autistic people.

Neural columns consist of a group of neurons that activate and eventually tune to specific stimuli. Neurons within the column do not have identical receptive fields and this allows for stimulus variability in the activation of the column. In this way, wider columns facilitate generalisation and narrow columns facilitate sensory discrimination. Gustafsson argued [27] that excessive lateral inhibition results in narrow neural columns. He further proposed that SOMs provide a biologically plausible way to model characteristics of autistic cortical maps.

Similarly to a brain cortical map that retains salient perceptual stimuli, a SOM can form representations of environmental input features and can exhibit analogous deficiencies to an autistic brain cortical map, if its formation mechanism becomes impaired.

**3.2 Modelling Autistic Traits**

The initial suggestion by Gustafsson [27] was that SOM impairment with biological plausibility entailed a lateral feedback inhibition increase of synaptic neuron weights during training, because of his claimed association of excessive lateral inhibition with narrow neural columns in the cortex. Later, he restricted this argument to the case of autism without co-morbid epilepsy [80] due to results presented by a subsequent study [81]. Excessive lateral inhibition can cause deficiencies in the generalisation capacity of a map and its feature representation, and establish narrow neural columns with almost identical activation patterns at the expense of variability. Such neural activity points to strong sensory discrimination and exaggerated feature specificity, even to the point of instability, and can result in the formation of partially developed or inadequate maps.

This modelling premise of an autistic type of impairment in SOM can be implemented by narrowing TN prematurely during the neural network training session. On IPSOM-A, the initial TN width ($\sigma_0$) parameter of the TN width function has a direct impact on the map's representational capacity aligned with Gustafsson's neural circuit theory [82]. In this case, a non-impaired cortical map would represent all the statistically dominant puzzle completion strategies with gradual transition between patterns. IPSOM-A can model such neural functioning using the original parameter configuration of IPSOM (i.e., including a typical initial size of the TN width).

In this study, we incorporated TN parameter modifications to IPSOM-A, and performed an evaluation. A series of four groups of 25 computer simulations per group were performed with the initial width of the TN width function set to a typical value of $\sigma_0=3$ (i.e., equal to the network's radius [33]; that is, half the number of neurons across in IPSOM's 6×6 map configuration) for one group, as originally used on IPSOM, and reduced to a narrow $\sigma_0=1.15$ for another simulation group. The latter value was selected heuristically, based on the map's size, as a relatively narrow one, although moderately narrow when compared to $\sigma_0$ values that are much closer to zero. Both simulation groups were performed twice, i.e., using a standard TN width function, in one simulation series, and an oscillating TN width function in a second one. For each SOM parameter configuration, the neural network was trained 25 times, each time being reset using a distinct SOM weight initialisation based on a different randomisation seed.

In the simulations, randomisation seeds were drawn from a sequence of computer pseudo-randomly generated numbers for the purpose of a non-biased SOM weight initialisation. A meta-seed was used to produce that sequence of pseudo-randomly generated numbers to ensure an impartial-root choice. This means is that the seeds were originally chosen as random, but the same random seeds were used for each variant of the algorithm, so that they all could be said to have started in the same 'random' state, so that none had an advantage derived from starting from a seed that happened to be a good one.

The results from the computer simulations confirmed that, for a large value of $\sigma_0$, the trained IPSOM-A forms an efficient representation of the input space (i.e., complete representation of the training set patterns with smooth transitional patterns present) whereas IPSOM-A training, using a small $\sigma_0$ value, results in a map with structural characteristics resembling autistic traits. Based on the SOM formation characteristics that were obtained, the simulation results also support the claim that the oscillating TN width is equivalent to the standard TN width IPSOM-A in modelling autistic traits. Both findings are discussed next.

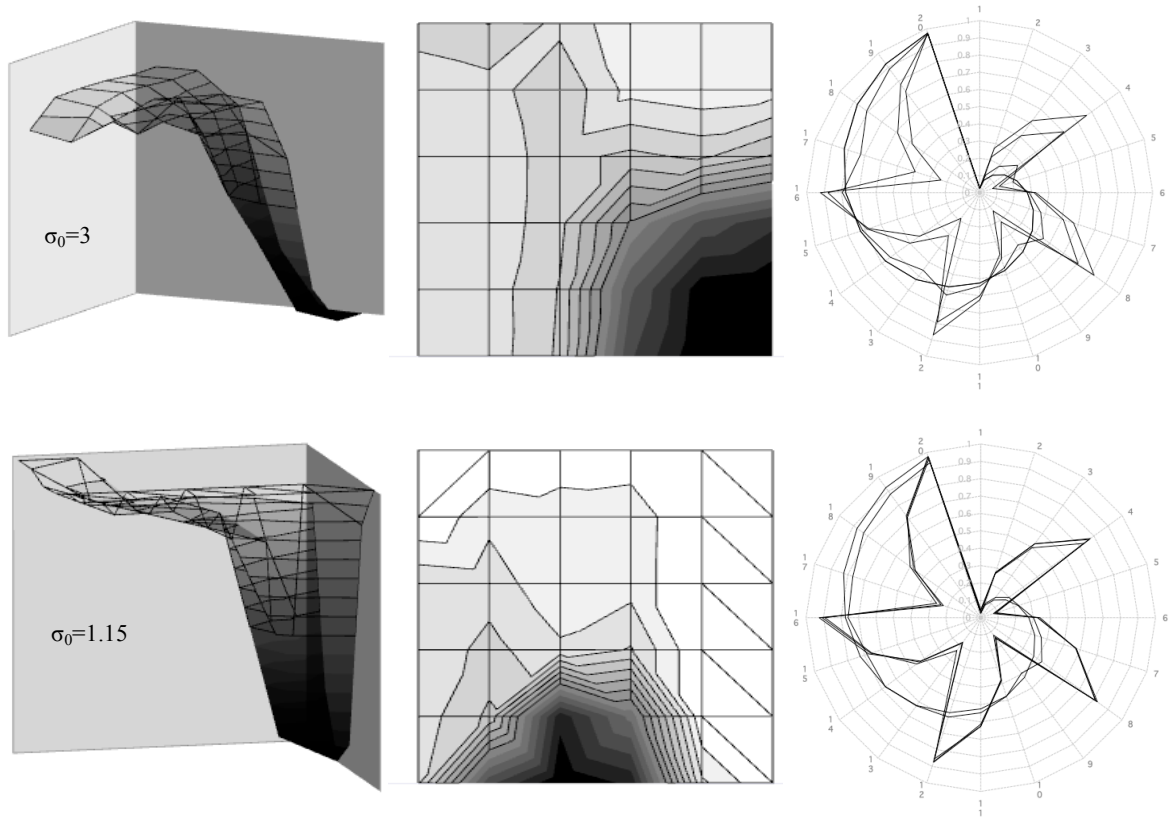

**Figure 7: Typical standard TN width IPSOM-A map characteristics. The significance of the three components for each value of $\sigma_0$ is explained in the text.**

Figure 7 depicts a characteristic example of trained IPSOM-A neurons from the simulations, for $\sigma_0=3$ (top three graphs) and $\sigma_0=1.15$ (bottom three graphs). The leftmost 3D graphs and their 2D versions in the middle show the Euclidean distance of each map neuron to pattern H (cf. Figure 5). The H pattern is one of the training set patterns and it is discussed here as an example. In both the 3D and 2D graphs, the darker areas indicate a smaller distance to H i.e., areas with higher representational accuracy for this pattern. The vertical axis on the 3D graphs also measures distance to pattern H: the higher the plot point from the base plane (the map) the larger the distance to H in the corresponding map area. Setting $\sigma_0=3$ results in a trained map with a more gradual transition between pattern H and other patterns, whereas setting $\sigma_0=1.15$ yields a map-wide steeper increase of the Euclidean distance signifying deficiencies in the representation of transitional patterns. This is shown on the 3D graphs by a steeper increase away from the base plane, whereas on the 2D graphs it is depicted as a tighter concentration of the darker areas.

The rightmost panels in Figure 7 contain superimposed radar graphs from the simulations of five neighbouring IPSOM-A map neurons for $\sigma_0=3$ (top) and $\sigma_0=1.15$ (bottom). Each radar graph depicts the resulting visualised overlaid patterns of five trained neurons; the patterns shown can be easily identified as H and V (cf. Figure 5) including transitional H and V patterns. Training IPSOM-A for $\sigma_0$=3 (top radar graph) results in a smoother (more gradual) transition between H and V patterns (i.e., more distinct transitional patterns), whereas for $\sigma_0$=1.15 (bottom radar graph) neurons are tightly grouped into two patterns (H and V) exhibiting a deficiency in pattern transition and generalisation: no transitional patterns are present, which is indicative of a high-discrimination map structure. Such a map will fail to identify variations of either pattern. Unlike the case of $\sigma_0$=3, it also contains redundant columnar ensembles (multiple H and V patterns) that distort the map's perceptual capacity: identical patterns, being represented by multiple neurons, will compete in firing during retrieval as if they were different [27].

In Figure 8, the same type of graphs depict a characteristic simulation example of a trained IPSOM-A (again, for $\sigma_0$=3 in the top graphs and $\sigma_0$=1.15 in the bottom graphs) where its standard TN width function during training was replaced with an oscillating TN width function as described in Section 2.2. The observations are very similar to those made for Figure 7. The differences between the centre panels of Figures 7 and 8 are due to the fact that pattern H was formed in different parts of the map in the two SOM training sessions. Also, the small difference between the radar graphs (right panels) of Figures 7 and 8, for $\sigma_0$=3, are due to formation variations of the particular simulations and do not imply a trend.

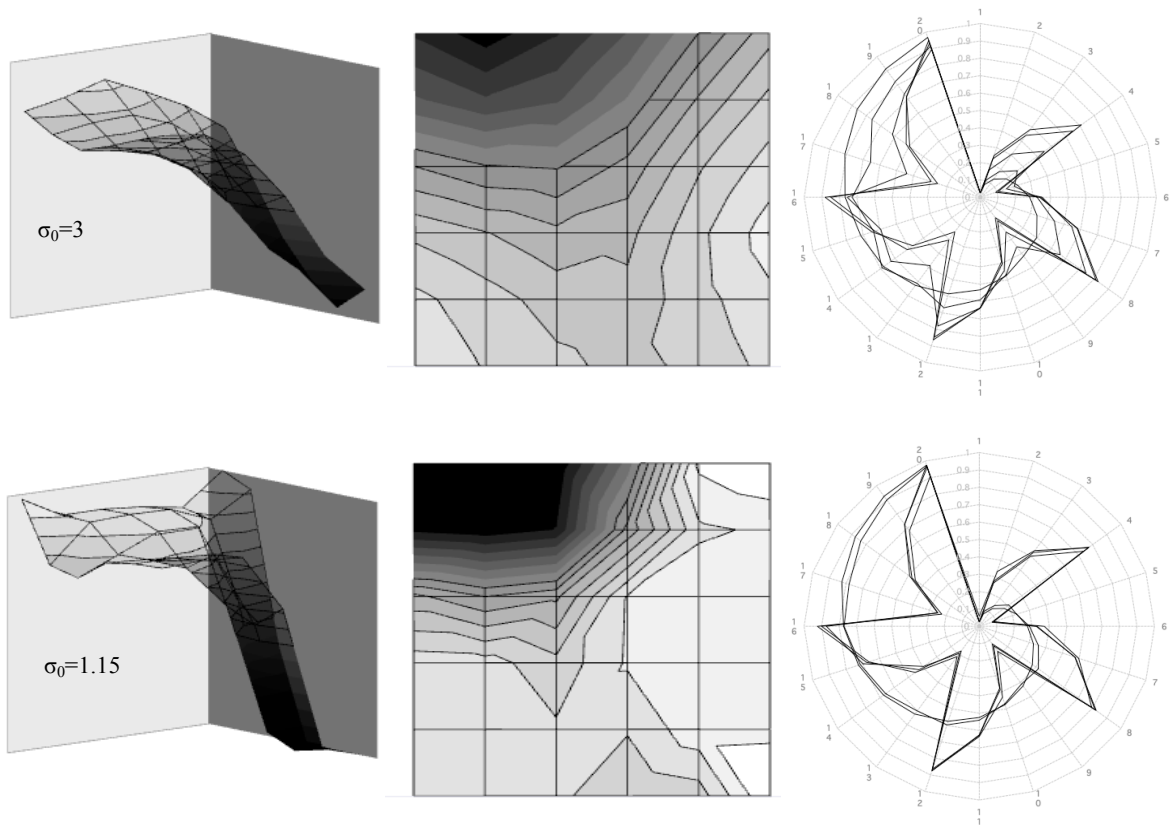

**Figure 8: Typical oscillating TN width IPSOM-A map characteristics. The significance of the three components for each value of $\sigma_0$ is explained in the text.**

The examples from the simulation results, illustrated in Figures 7 and 8, are representative in terms of the observed characteristics. The specific patterns (H and V) shown in the radar graphs were chosen in order to demonstrate clearer the map's transitional behaviour since they exhibit a low correlation significance (Table 1) compared to the other training patterns.

**Table 1: Correlation between IPSOM training patterns.**

| Spearman's ρ *(N=20)* | | H | V | PH | PV | Spearman's ρ *(N=20)* | | H | V | PH | PV |
|---|---|---|---|---|---|---|---|---|---|---|---|
| Correlation Coefficient | H | 1 | .429 | .523* | .507* | Significance (2-tailed) | H | . | .059 | .018 | .023 |
| Correlation Coefficient | V | | 1 | .388 | .420 | Significance (2-tailed) | V | | . | .091 | .066 |
| Correlation Coefficient | PH | | | 1 | .974# | Significance. (2-tailed) | PH | | | . | .000 |
| Correlation Coefficient | PV | | | | 1 | Significance (2-tailed) | PV | | | | . |

Correlation is significant at the 0.05 level (*) and at the 0.01 level (#).

# 4. The IPSOM-D Model

## 4.1 A Neurocomputational View of Delusions in Schizophrenia

The research on schizophrenia has been continuous, spanning a century, and has yielded an ever-changing understanding of this complex disorder [71][72][73][74][83]. Today, schizophrenia is widely regarded as a progressive neurodevelopmental disorder [84][85][86][87][88][89]; one of its most typical psychotic positive symptoms is delusions [90].

Spitzer has argued [91] that acute delusions are not "false statements" but statements lacking appropriate context and justification; they are overemphasised beyond their actual significance. Although acute delusions can be seen as a state or mode of operation, repetition or prolongation of it eventually incorporates them into the cortical map structure as part of the person's 'belief' system and they become chronic. He presented a modelling claim [91][22][38], according to which SOM artificial neural networks could be used as a model of brain cortical function, implementing a key neural process - lateral inhibition. Spitzer built a neurocomputational explanatory framework for delusions around the properties of brain neuroplasticity and neuromodulation in relation to the formation and operation of sensory and higher-order (i.e., relating to more complex semantic information and to concept hierarchies) computational cortical maps.

Specifically, neuromodulation is associated with the information-theoretic view of neuron signal-to-noise ratio. Noise is an essential ingredient of change in cortical representations [92]. In a high signal-to-noise ratio mode of operation, memory retrieval becomes more reliable but change becomes unlikely. High neuromodulator activity, with increased noise-to-signal ratio, intensifies focusing in neuronal activation, corresponding to narrow cortical columns, and is linked to acute delusional states; this type of focusing can be modelled in a SOM neural network by implementing excessive lateral inhibition. Facilitated by brain neuroplasticity, temporally persistent acute delusional states can establish entrenched (dominant) and distorted higher-order cortical maps; this process can be associated with the formation of chronic delusions. Once chronic delusions have been established, it becomes increasingly difficult to revert to a non-delusional stage. Again, this is consistent with SOM network behaviour: formed maps require complete and prolonged re-training, using a reduced lateral inhibition mode, to alter their representations effectively.

Furthermore, it has been proposed [93][94][95][96] that a principal phenotype of schizophrenic delusions includes spurious or parasitic attractors at the cognitive level. Spurious or parasitic attractors involve the creation of autonomous memory patterns that do not correspond to environmental stimuli; such patterns can directly lead to the formation of delusions. Chen and Berrios [96] note as two of their global factors in delusion formation a) any bias in favour of internal representations due to e.g., perception impairment, and b) a neural imbalance of signal-to-noise ratio, where over-amplified signals result in cognitive inflexibility including resilience to memory revision and environmental change. Although Chen and Berrios' and Hoffman and McGlashan's work on attractors does not share the SOM approach, it is nevertheless related at the neuroscientific level, in the sense that it refers to a common functional cortical set of characteristics and resulting behaviour (i.e., creation of autonomous non-environmental memory patterns, bias in favour of internal representations, and imbalance of increased signal-to-noise ratio).

Neurocomputational approaches to brain disorders are valuable in the context of cognitive modelling. Theoretical proposals employing such an approach, including Gustafsson's and Spitzer's, often involve analogues or correspondence (typically, not a complete equivalence) between computational and biological mechanisms, as well as the indication of a link between the latter and the phenotype of a disorder. A common computational mechanism, as is the case here with SOM TNs, employed on two different brain disorders, does not implicitly or explicitly equate the disorders. As mentioned earlier, SOM networks provide a computational tool that is suitable to model core biological aspects of brain function and organisation, and TN is central in a number of related theories, including Gustafsson's and Spitzer's. However, any correspondence is strictly with certain important biological mechanisms that, in these two cases, are, in turn, claimed to be linked to elements of the disorder's phenotype set. This line of cognitive modelling argumentation is not 'designed' to justify or imply coverage of the entire spectrum of symptoms and causal mechanisms for the disorder. We will return to this point later.

### 4.2 Modelling Key Elements of Acute and Chronic Delusions

The degree of neuromodulation in the cortex plays a significant role in the clinical manifestation of acute delusions [91]; this impacts the signal-to-noise ratio modulation. From a SOM modelling perspective, control of the lateral inhibition can be considered an analogue for signal-to-noise ratio control that is linked to focusing in neuronal activation (i.e., the size of neural columnar ensembles). In IPSOM-D this is implemented by regulating the width of TN during training. The modelling hypothesis for IPSOM-D is that the initial TN width ($\sigma_0$) parameter, part of the TN width function, can regulate TN during map formation as discussed. Consequently, a premature reduction of TN could 'induce' the representation of patterns in the map consistent with characteristics of acute delusions.

As with IPSOM-A, TN parameter modifications to IPSOM-D were incorporated, and an evaluation was made. A series of two groups of 25 controlled simulations per group were performed: the initial TN width was set to $\sigma_0=3$ (identical to that of IPSOM) for the first group, and adjusted to a narrow $\sigma_0=1.15$ for the second group. A third smaller group also was added using a further narrowed down value of $\sigma_0=0.7$. These values were heuristically selected, based on the size of the map, to cover the entire rage of initial TN width values, without reaching extremes that would essentially prevent SOM training altogether. The computer SOM training simulations were performed for all three groups using: a) a standard TN width function, and b) an oscillating TN width function. As with IPSOM-A, initial SOM weight randomisation was employed, and the neural network was reset and re-trained on each simulation per group, each time using a distinct SOM weight initialisation based on a different randomisation seed.

The simulation results show that IPSOM-D has the capacity to represent the input space efficiently when $\sigma_0$ is large. However, for small $\sigma_0$ values, the resulting map contains atypical structures that can be interpreted as corresponding to chronic delusions. Based on the obtained SOM formation characteristics, the results also support a 'modelling equivalence' claim between the oscillating TN width and standard TN width versions of IPSOM-D. Both findings are discussed next.

Dominant or entrenched structures that could induce a chronic delusion phenotype can be observed by examining map distortions in a trained IPSOM-D (i.e., a map failing to represent the entirety of input space, containing too many transitional patterns, or forming non-existent patterns) from a given untrained (initial) state. In this context, a 'delusional' structure can be identified as a group of trained neurons that represent a pattern either dissimilar to any training or transitional pattern, or excessively dominant (or under-represented) in the map; the latter case compromises the SOM property of density matching [34], i.e., the representation of the input space statistical distribution. In addition, resistance to representing patterns can be an interpretational indicator of established (chronic) 'delusional' structures [91][96].

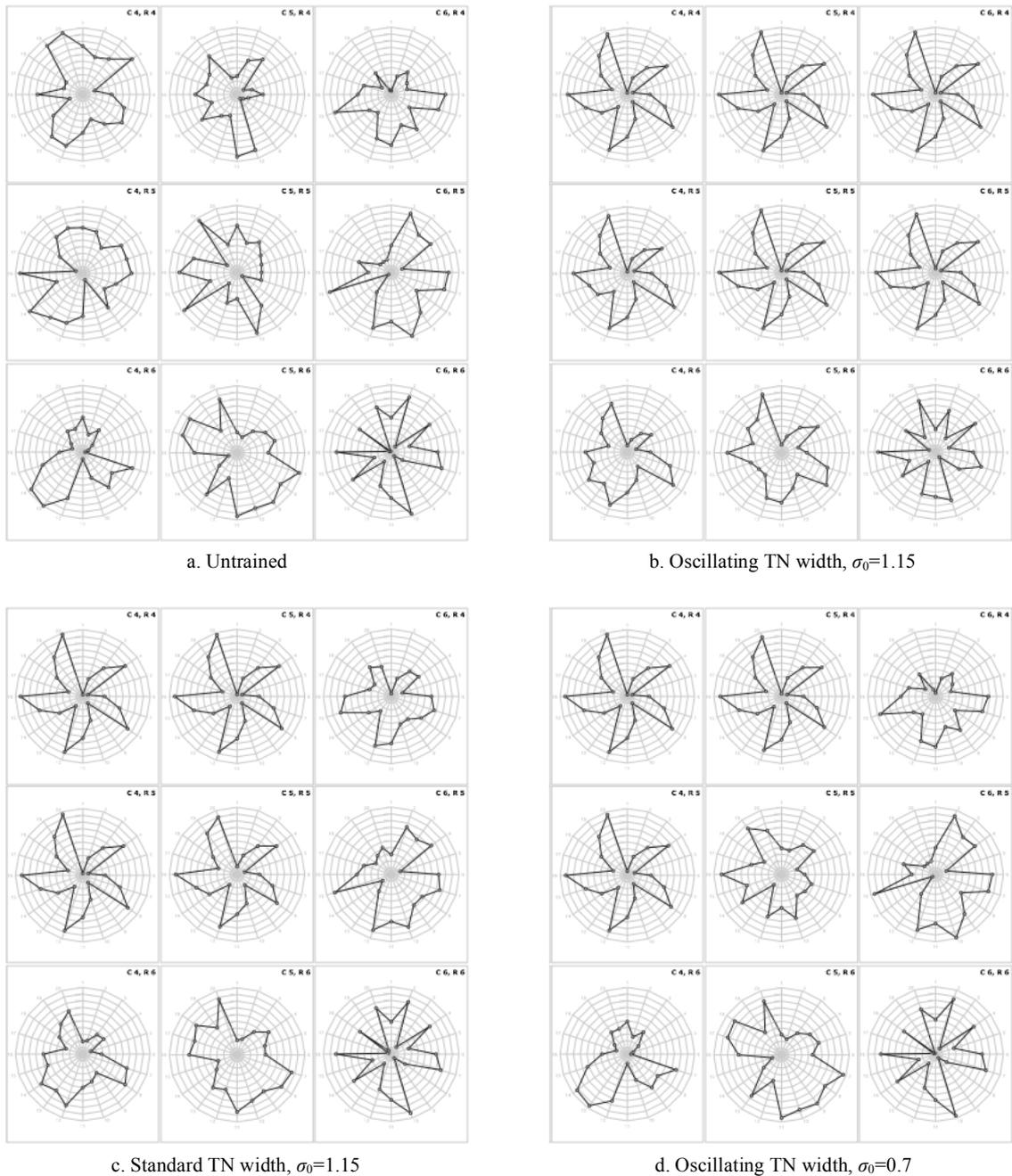

a. Untrained        b. Oscillating TN width, $\sigma_0$=1.15

c. Standard TN width, $\sigma_0$=1.15        d. Oscillating TN width, $\sigma_0$=0.7

**Figure 9: Induced delusional structures on IPSOM-D.**

Four snapshots (Figure 9) of the bottom-right IPSOM-D quadrant depict the untrained state (panel 9.a) together with three different map formation outcomes (panels 9.b, 9.c, 9.d) for different SOM configurations. Each panel contains nine visualised IPSOM-D patterns in the form of radar graphs, corresponding to nine map neurons. In panel 9.a we see the situation before training (i.e., random initial patterns); panels 9.b, 9.c, and 9.c depict the same map quadrant after training for different small $\sigma_0$ value and TN width function configurations. In Figure 10, as a reference, the same quadrant of the IPSOM-D map is shown for a typical value of $\sigma_0=3$ using a standard TN width function. A first observation after comparing panel 9.a (initial state) with panel 9.c (trained map using a standard TN width) is the perseverance of numerous initial pre-training patterns indicating impairment in the acquisition of environmental (in this case, training) stimuli: over half of the quadrant's neurons represent either the original (initial) random pattern or a distorted (parasitic) version of it. In panel 9.b (trained map using an oscillating TN width) there is an additional observation of an excessive representation of the V pattern (cf. Figure 5). Once the $\sigma_0$ becomes particularly narrow (panel 9.d) the observed 'delusional' indicators become exceedingly prominent: resistance to environmental change by retaining the original or a distorted version of the columnar configuration, and the occurrence of parasitic patterns (i.e., patterns unrelated to environmental stimuli).

Section 5.2 further discusses various cognitive modelling aspects and parameters in regard to delusions in schizophrenia (as well as autism, presented in Section 3). It also poses a question relating to any potential similarities at the functional or cortical map level between the two disorders, as implied by the SOM modelling approach and results presented in Sections 3 and 4.

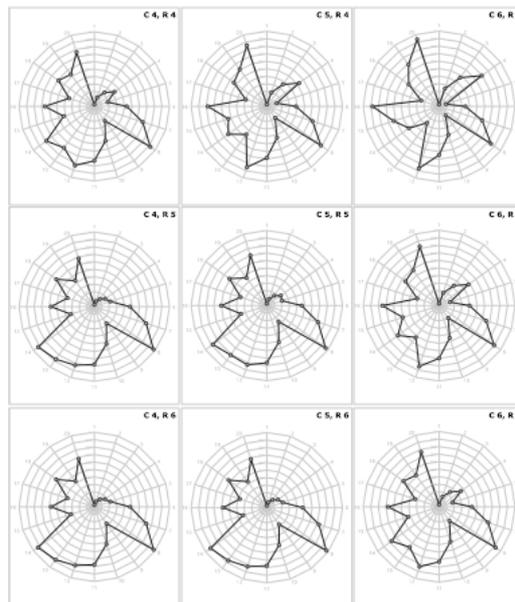

**Figure 10: Typical structure (Standard TN width, $\sigma_0$=3) on IPSOM-D.**

## 5. Conclusion

### 5.1 TN width area

There are biological and computational implications of the role of TN in SOM cognitive modelling. In this study, we introduced a modified TN width function with increased biological plausibility, and conducted computational modelling simulations based on two neurocomputational theories of respective neurodevelopmental disorders (autism, schizophrenia). As shown in the preceding sections, the modelling significance of the oscillating TN width function is associated with the initial TN width ($\sigma_0$) parameter. At a closer examination, this parameter has a decisive role in determining the area under the TN width curve through the epochs of SOM training - this area measures training 'opportunity'. From this perspective, what is considered 'narrow' or 'wide' TN during SOM formation is a function of the TN width area covered. In Figure 11, plots of the standard and the oscillating TN width functions are overlaid in two graphs. The TN width area is bordered above by the respective TN width function, and below by the epoch (horizontal) axis.

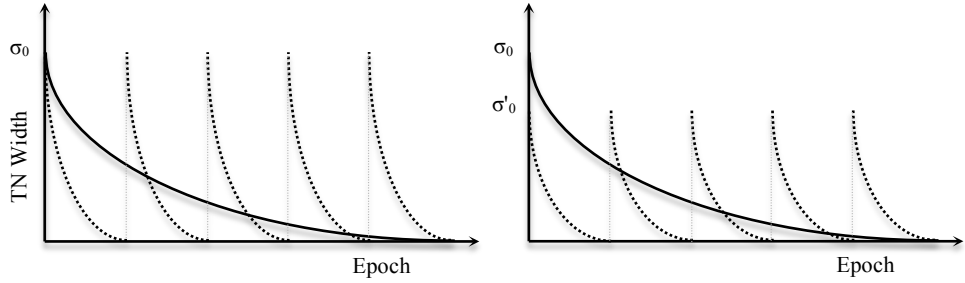

**Figure 11: Standard and oscillating TN width areas, for a typical $\sigma_0$ (left panel) and a reduced $\sigma'_0$ (right panel).**

Mathematically, because

$$\int \sigma_0 \cdot e^{\left(-\frac{x}{\tau}\right)} dx = \sigma_0 \cdot (-\tau) \cdot e^{\left(-\frac{x}{\tau}\right)} + C, \quad \sigma_0, \tau \in R$$

we can calculate the area for a specific TN width function, initial TN width $\sigma_0$, and number of epochs $t$, using the following formula:

$$\text{Area under } \sigma(x) \text{ curve} = \int_0^t \sigma_0 \cdot e^{\left(-\frac{x}{\tau}\right)} dx = -\left[\sigma_0 \cdot \tau \cdot e^{-\frac{x}{\tau}}\right]_0^t = -\sigma_0 \cdot \tau \cdot \left(1 - \frac{1}{e^{\left(\frac{t}{\tau}\right)}}\right)$$

In the standard and oscillating TN width simulation results for IPSOM-A and IPSOM-D, the calculated value of the *area under $\sigma(x)$ curve* (for the same value of $\sigma_0$) remained unchanged irrespective of the TN width function used. Thus, the two approaches (i.e., standard and oscillating TN width) provide the 'same amount' of σ to the training process. This verifies the output equivalence between the two modelling approaches. Furthermore, in the oscillating TN width function simulations, when the $\sigma_0$ value was reduced to $\sigma'_0$, the calculated value of the *area under $\sigma(x)$ curve* was significantly smaller (Figure 11, right panel) and resulted in an IPSOM-D map with more pronounced 'delusional' structures (Figure 9, panel 9.d). This demonstrates the computational and cognitive modelling significance of the *area under* the TN width *curve* and its relationship to TN width.

### 5.2 Discussion and Future Work

It is reported in the literature that although SOM neural networks are relatively uncomplicated to implement, they strongly resist formal analysis [97]. Making a link between the biological, behavioural and computational levels in cognitive modelling studies often requires a sequence of finely drawn associations across disparate disciplines [98]. However indirect and interdisciplinary such a link may be, the methodology and tools to create it have long been available, and an effort was made in this paper to construct it.

In summary, this study: a) *demonstrated a SOM computational validation of certain neurocomputational theories of neurodevelopmental disorders, specifically, autism and schizophrenic delusions*, b) *introduced an oscillating TN modified SOM with increased biological plausibility (biological realism),* and c) *postulated the cognitive modelling significance of the initial TN width parameter and the TN width area in the general SOM TN modelling hypothesis.*

The computational validation of the neurocomputational theories discussed in Sections 3 and 4 has been limited to focusing on certain aspects of each disorder under examination. Gustafsson's autistic theory, for instance, focuses on perceptual impairments rather than deficits in socialisation or linguistic communication. To our knowledge, no computer simulation study, directly related to either theory, has been published. The use of a carefully designed complex weight-vector-encoding computational model to validate these theories is an incremental step towards demonstrating the potential of SOM cognitive models in psychiatry and clinical neuropsychology. Irrespective of the conclusive neuroscientific validity per se of the theories examined, another principal goal was to demonstrate how different levels of explanation (behavioural, biological, and computational) can be linked to provide a complete, constructive and unified result for all disciplines involved (including psychology, psychiatry, neuroscience, and computer science), and potentially lead towards therapeutic/diagnostic implications for such overwhelming and resistant-to-treatment mental disorders in the future. As an example, potential medical insights could be obtained from the observation of the nature and temporal architecture of the SOM pattern distortions. The latter could be used to base new working hypotheses for novel methods of detecting atypical brain disorder phenotypes; a somewhat similar approach has been suggested in the case of Hopfield networks for Alzheimer diagnosis [99].

A very recent work on SOM brain disorder modelling focuses on semantic dementia [100]; it employs a combination of biased SOM training and a form of neuronal amputation, both of which modifications are consistent with a neurodegenerative disorder rather than the different approach followed in this study of modelling neurodevelopmental clinical phenomena via atypical formation. The importance of distinguishing between the two types of disorders can be exemplified by noting here another very recent work [101], a large-scale longitudinal study examining cognitive decline/dementia and autism: the study found no significant evidence that higher autistic traits in middle-aged and older adults either negatively or positively affected their spatial working memory over a period of seven years. (This result focuses on the trajectory of decline, rather than the more widely recognized impact of autism on spatial working memory.)

There does not seem to be any relevant study that uses time-varying SOM networks for behavioural or neurodevelopmental modelling (see [102]) for rather different time-varying SOM networks for robotics applications); so SOM TN oscillation is novel in this regard. The introduction in this paper of an oscillating TN modified SOM with increased biological plausibility has the potential to improve future cognitive modelling work using SOM neural networks, especially in the fields of computational psychiatry and clinical neuropsychology. In the case of schizophrenic delusions, a single-descent TN width SOM model would intuitively present a less plausible computational analogue of the formation of acute delusions because it lacks the dynamics of a fluctuating neuron activation width that is required in order to produce the repeated narrow-columnar activity suggested by the disorder's phenotype. The multi-level neural oscillation observed in the brain calls for further examination of its potential correspondence with computational SOM mechanisms at the biological, functional and behavioural levels.

The neuroconstructivist approach to cognitive development [103][104] stresses the importance of the initial conditions, at the neurobiological level, in determining atypical developmental trajectories and the emergence of a developmental disorder's phenotype. This study underlined the increased cognitive modelling significance of the initial TN width parameter (and the associated TN width area) in the general SOM TN modelling hypothesis. However, the significance of synaptic initialisation of a SOM neural network also needs to be examined more closely. The conventional method of random-weight initialisation provides a computationally non-biased approach but it might be detracting from constructing a universal biologically realistic model. Although, prenatally formed ontogenetic cortical neural columns [105] are not the result of environmental interaction, they cannot accurately considered to be 'random' in the way a SOM initialisation procedure would imply. In certain cases, where the impact of this procedure is dissolved in the SOM training process (i.e., does not affect the SOM formation in a detectable way in regard to the particular cognitive model being studied), the question of biological significance can be bypassed. A common alternative method is to initialise the network using a number of vector inputs, drawn at random, from the training set. It remains to be investigated, however, whether such a choice could suggest a biologically plausible innateness-equivalent for a cortical map, given the fact that it originates from the environment (as the training set typically corresponds to) rather than of a more genetically oriented 'internal' source.

DSM and ICD classifications [71][72][73] fail by design to properly address the biological dimension of mental disorders [106]. A different research project (RDoC) has emerged that aims to examine abnormal brain functioning from a different standpoint, inclusive of biological mechanisms and computational approaches, that crosses the boundaries between currently classified disorders and eventually redefines them under a new diagnostic framework aiming to inform future diagnostic systems [107][108][109]. In this study we examined the associations between biological structures and mechanisms, behavioural traits, and biologically plausible computational structures and mechanisms of two neurodevelopmental disorders that are currently classified as entirely different disorders. It remains an open question to what extent this classification remains relevant at the level of causal neurodevelopmental mechanisms.

The similarities between IPSOM-A and IPSOM-D are explicit at the computational level and can be measured quantitatively. As discussed in Sections 2.1, 3, and 4, the key computational element, which is shared between the two models, is the SOM TN width and is directly affected by the initial TN width parameter. From a cognitive modelling standpoint, the TN width implements the control of cortical neural columns in terms of size, and regulates the amount of SOM lateral inhibition. In both neurocomputational theories, the formation of narrow neural columnar ensembles plays a decisive role.

At the same time, the neurobiological phenomena being modelled belong to two disorders that are considered very distinct, and each disorder displays a spectrum of neuropsychological and psychiatric behaviours that do not appear to be very compatible. This fact rather masks any potential implications of such similarities for the two disorders. However, the gap between the modelling levels (computational, neuroscientific, behavioural) as well as between their type (functional, biological, psychological) allows for valid claims to be made. Specifically, if IPSOM-A and IPSOM-D demonstrate at a functional level (whether computational or neuroscientific) certain common neurodevelopmental mechanisms and/or structures, and support distinct neuropsychological or psychiatric theories, then that in itself is a valid claim to be considered at the level that it is made. It is certainly conceivable that similarities at this level can exist without the requirement of obvious corresponding wide span similarities in phenotypes. Furthermore, whether common behavioural aspects between schizophrenia and autism (or, as it is more accurately referred to today, "Autism Spectrum Disorder", ASD) exist could be partially a matter of classification. Last but not least, a certain connection between the two disorders has long been acknowledged and currently described in terms of schizophrenia negative symptoms partial correspondence with ASD social deficits, co-morbidity patterns between the two disorders, and shared cognitive impairments as well as potential common biological underpinnings that may begin very early in neural development [110][111][112][113]. Under the light of current neurocomputational modelling this connection may be shown to be even more substantial than originally considered.